\documentstyle[emulateapj, epsfig,psfig]{article}

\begin{document}

\title{Tidal stirring and the origin of dwarf spheroidals in the Local Group}

\author{Lucio Mayer $^{1,4}$, Fabio Governato $^2$, 
Monica Colpi $^1$, Ben Moore $^3$,
Thomas Quinn$^4$, James Wadsley$^{4.5}$, Joachim Stadel$^4$ \& 
George Lake$^4$}

\affil{$^1$Dipartimento di Fisica, Universit\`a Degli Studi di
Milano Bicocca, via Celoria 16, I--20133 Milano, Italy\\
$^2$ Osservatorio Astronomico di Brera,
via Bianchi 46, I--23807 Merate (LC) - Italy\\     
$^3$ Department of Astronomy, University of Durham, Durham, U.K, DH1 3LE\\
$^4$ Department of Astronomy, University of Washington
Seattle, USA, WA 98196\\
$^5$ Department of Physics and Astronomy, McMaster University,
Hamilton, Ontario L8S 4M1 Canada}

\begin{abstract}

N-body + SPH simulations are used to study the evolution of dwarf
irregular galaxies (dIrrs)  entering the dark matter
halo of the Milky Way or M31 on plunging orbits. 
We propose a new dynamical mechanism
driving  the evolution of gas rich, rotationally supported
dIrrs, mostly found at the outskirts of the Local Group (LG), 
into gas free, pressure supported dwarf spheroidals (dSphs) or 
dwarf ellipticals (dEs), observed to cluster around 
the two giant spirals.
The initial model galaxies are exponential disks embedded 
in massive dark matter halos and reproduce nearby dIrrs.
Repeated tidal shocks at the pericenter of their orbit partially 
strip their halo and disk and trigger dynamical instabilities 
that dramatically reshape their stellar component.
After only 2-3 orbits low surface brightness (LSB) dIrrs 
are transformed into dSphs, while high surface brightness (HSB) dIrrs
evolve into dEs. 
This evolutionary mechanism naturally leads to the morphology-density
relation observed for LG dwarfs. 
Dwarfs surrounded by very dense
dark matter halos, like the archetypical dIrr GR8, are turned into
Draco or Ursa Minor, the faintest and most dark matter dominated among
LG dSphs. 
If disks include a gaseous component, this is both tidally
stripped and consumed in periodic bursts of star formation.
The resulting
star formation histories are in good qualitative agreement with those
derived using HST color-magnitude diagrams for local dSphs.


\end{abstract}

\section{INTRODUCTION}

Dwarf galaxies in the Local Group (LG) clearly obey a morphology-density
relation.  Close to the Milky Way and M31 we find early-type dwarf
galaxies, namely faint ($M_B > -14$) low surface brightness dwarf spheroidals
(dSphs) and more luminous ($M_B > -17$), higher surface brightness
dwarf ellipticals (dEs). All these galaxies are nearly
devoid of gas, contain dark matter and mainly old stars
and are supported by  velocity dispersion (Ferguson \& Binggeli 1994, 
hereafter FB94; Mateo 1998, hereafter Ma98;
Grebel 1999, hereafter Gr99; Van den Bergh 1999).  
Among them Draco and Ursa Minor have the highest dark matter densities 
ever measured (Lake 1990). On the outskirts of the LG we find 
similarly faint ($M_B>$ -18) and dark matter dominated dwarf irregular
galaxies (dIrrs), that are gas rich, star-forming systems with disk-like kinematics (Ma98, Van den Bergh 1999, Gr99).

Previous attempts to explain the origin of dSphs in the 
LG have relied on gas dynamical processes to remove the gas in
dIrrs.  Gas stripping may result either because of the pressure
exerted by an external hot gaseous medium in the halo of the Milky Way
(``ram pressure'') (Einasto et al. 1974) or because of internal strong
supernovae winds (Dekel \& Silk 1986).  However, ram pressure would
require an external gas density that is several orders of magnitude
higher than recently inferred for the Milky Way (Murali 2000) and
supernovae winds cannot explain the existing
morphology-density relation.  Moreover, such dissipative mechanisms
would remove the gas but would not directly alter the structure 
and kinematics of the pre-existing stellar component.
However, the light follows an exponential profile in both dSphs and dIrrs
(Faber \& Lin 1983; Irwin \& Hatzdimitriou 1995; Ma98)
and a positive correlation between surface brightness and luminosity
is shown by both types of dwarfs (FB94), suggesting an evolutionary link
between them.
Is there a mechanism that can transform dwarf galaxies between 
morphological classes or must we
accept the idea that dSphs are fundamentally different from dIrrs ?

Within rich galaxy clusters, fast fly-by encounters with the largest
galaxies can transform a disk system into a spheroidal or S0 galaxy in
just 3-4 Gyr (Moore et al. 1996, 1998).  If the halos of bright
galaxies were scaled down versions of galaxy clusters then this
``galaxy harassment'' would be equally important within them.
However, whereas rich clusters contain over thirty large ($L_*$)
perturbing galaxies, the Milky Way and M31 have only a couple of
satellites sufficiently massive to harass the other dwarf galaxies
(Moore et al. 1999; Klypin et al. 1999) As a result, the rate for
effective satellite-satellite fly-by encounters is less than one in
every 10 Gyr (the LMC and the SMC being a notable exception).

Thus, we are left only with the repeated action of tidal forces from
the primary galaxy as an evolutionary driver. These operate on the
orbital timescale, which is of order of 3-4 Gyr in both clusters and
galactic halos.  However, given the relatively low age of large,
virialized clusters, galaxies have typically approached the cluster
center only once by the present time, while  dSphs satellites
have had sufficient time to complete several close tidal encounters
with the Milky Way, as stellar ages imply that the latter was already
in place  10 Gyr ago (Van den Bergh 1996).

In this paper we use  very high resolution N-Body + SPH 
simulations performed with the parallel binary treecode GASOLINE
(Dikaiakos \& Stadel 1996; Wadsley et al. 2000) to follow the evolution of small galaxies 
resembling dIrrs as they move on bound orbits in the tidal field of 
the massive dark matter halo of the Milky Way.

\section{Models of dwarf galaxies}

The Milky Way halo is modeled as the fixed potential of a truncated
isothermal sphere with a total mass $4 \times 10^{12} M_{\odot}$
inside a radius of $400$ kpc, consistent with both recent measures
based on radial velocities of distant satellites (Wilkinson \& Evans
1999) and with generic models of structure formation (Peebles et
al. 1989).  The core radius is $4$ kpc and the resulting circular
velocity at the solar radius is $220$ km/s.

\medskip
\epsfxsize=8truecm
\epsfbox{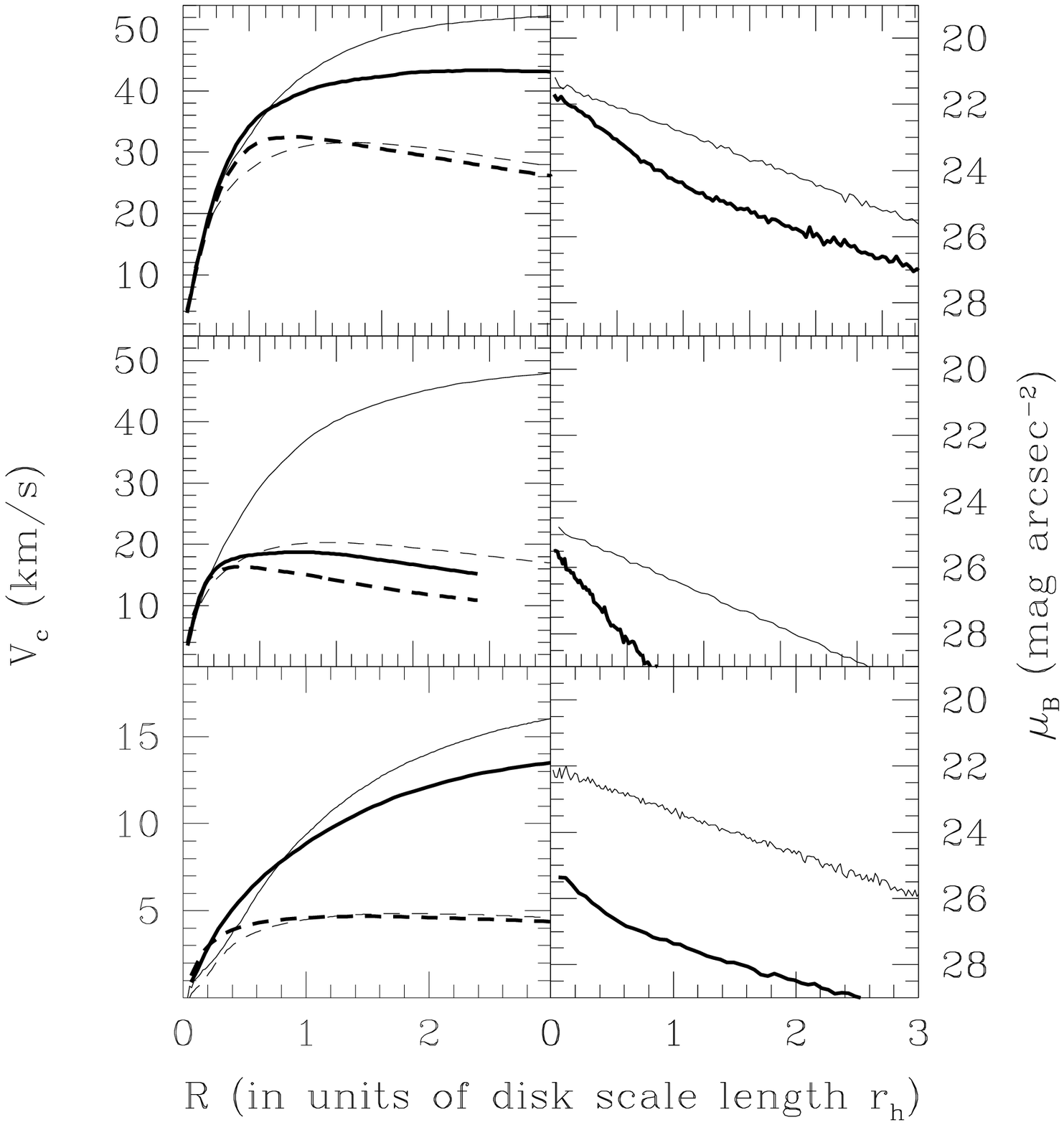}
\figcaption[panelprof.ps]{\label{fig:asymptotic}    
\small{Evolution of circular velocity (left) and surface brightness
profiles (right) for our model galaxies: thin lines are
used for the initial profiles, thick lines for the profiles
after 7 Gyr. From top to bottom, models $M2H$, $M2La$ and $GR8$
are shown.
The overall (stars + dark matter) circular velocity profile is 
always represented
by a solid line, while a dashed line
describes the 
contribution of the stellar component alone. 
For the final surface brightness profile we take into
account fading according to the star formation history described in the text.
We assume that the last burst occurs 4 Gyr ago for HSBs and LSBs and
7 Gyr ago for GR8, thus resulting in a more pronounced fading for the
latter. The different choices correspond to the different epochs of infall
inferred from our evolutionary model.
The orbits had an
apo/peri of 9 and a pericenter of $\sim 40$ kpc (HSBs and LSBs) and 12
kpc (GR8), with random disk orientations.}}
\medskip

Our simulated galaxies are modeled 
as exponential disks of stars with a Toomre parameter
$Q=2$ embedded in truncated isothermal dark matter halos (see Hernquist 1993).
The number of particles in the disk is 50,000 while that in the halo 
ranges from 250,000 to $3\times 10^6$. 
Such high resolution in the halo 
reduces considerably numerical heating of the disk due to massive
halo particles, as was tested by evolving our models in isolation for
5 Gyr.  The scale-lengths and masses of the stellar disk and dark
matter halo are chosen so that the satellites follow the observed B
band Tully-Fisher relation (Hoffman et
al. 1996; Zwaan et al. 1997) and have realistic rotation curves (de
Blok \& McGaugh 1997; Cote et al. 1997), as shown in Figure 1.
The determination of general scaling properties of galaxies becomes
more uncertain at the very faint end.
Current structure formation models predict that the mass $M$ and scale radius
$R$ of halos (and of their embedded disks) vary with redshift as
$\sim (1 + z)^{-{3/2}}$ for a fixed value of the circular velocity $V_c$,
with $V_c \sim R \sim M^{1/3}$ 
(Mo et al. 1998). 

\medskip
\epsfxsize=8truecm
\epsfbox{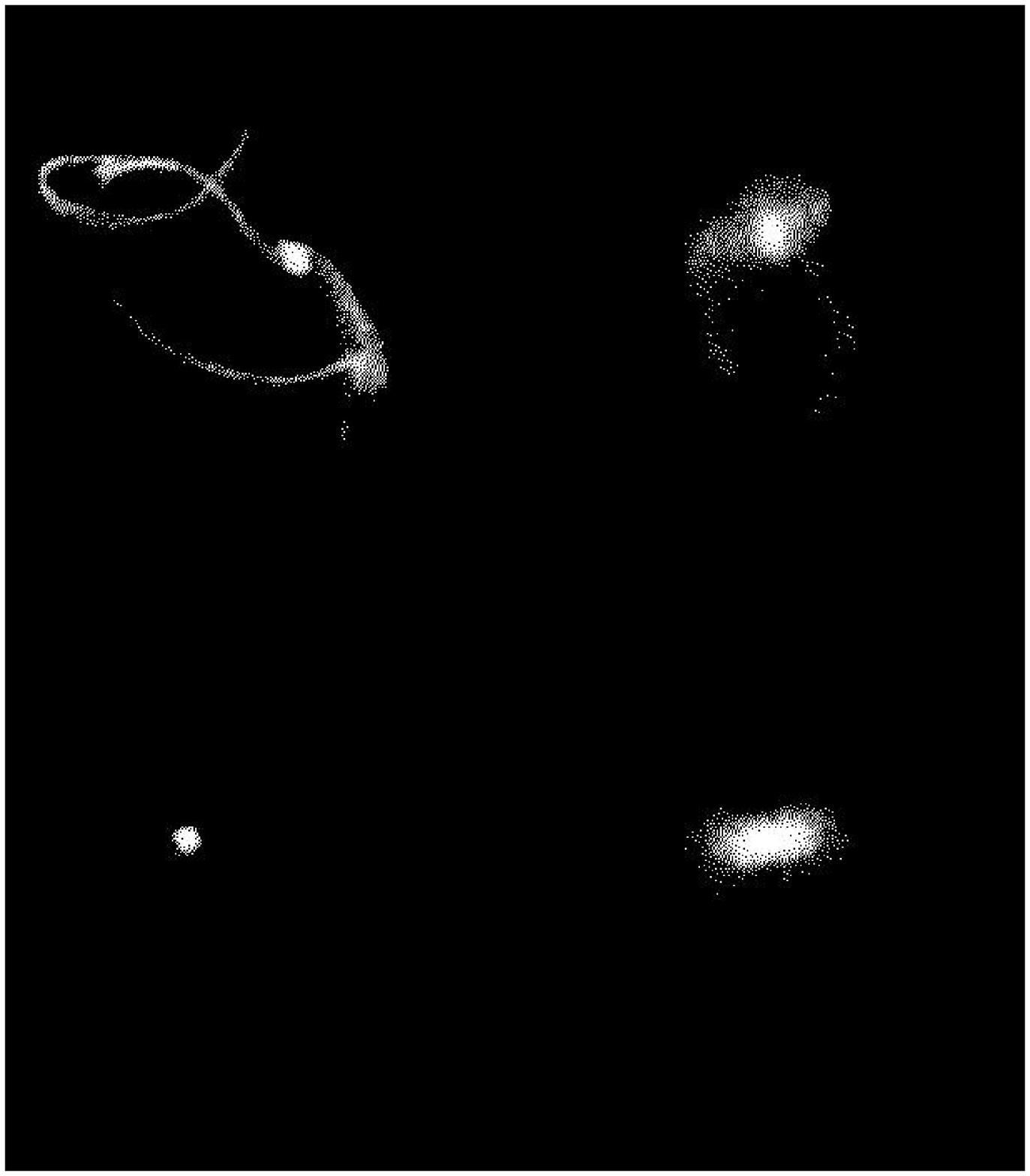}
\figcaption[panelscr2.ps]{\label{fig:asymptotic} 
\small{The final stellar configurations of our model galaxies.
Left panels show the stellar 
streams of the LSB (upper) and HSB (lower) satellites viewed face on
(the orbit has apo/peri=4).
The boxes 
are
500 kpc on a side, with brighter colors showing regions with higher
density.  
Right panels show close up views of the remnants seen edge-on
in frames of 50 kpc on a side.}}
\medskip

The scaling with redshift reflects the fact that low mass
halos form at early epochs and have higher characteristic densities
because the average density of the Universe was higher.
Galaxies with total masses $<$ 10$^9$ $M_{\odot}$ should typically 
form at $z \ge 2$ (Lacey \& Cole 1993)
and should live in halos with a central dark matter density 
$\sim$ 0.3 $M_{\odot}\rm {pc^{-3}}$, comparable to what inferred for GR8, an
extremely faint LG dIrr 
(Carignan et al. 1990). The model with the smallest mass (``GR8'')
was built following these prescriptions.

The models cover the entire luminosity function of irregulars in the
LG (Ma98) with stellar masses of $1.2 \times 10^6
M_{\odot}$ ($M_B$ = -11.2, ``GR8''), $9 \times 10^8 M_{\odot}$
($M_B$=-16.23, ``M2'' ) and $2.5 \times 10^9$ ($M_B$=-18, ``M1'').  
At fixed disk mass (either M1 or M2) we also vary the disk
scale length, obtaining high surface brightness (HSBs, $\mu_B=
21.5$ mag arcsec$^{-2}$, M1H and M2H) or low surface brightness
satellites (LSBs, $\mu_B= 24.5$  or $23.5$ mag arcsec$^{-2}$,
M1La, M2La, M2Lb) for a total of six different models (we always
assume $M/L_B$=2 for the stellar disk and set the halo core radius
equal to the disk scale length; see de Blok \& McGaugh 1997). 
The GR8 model was obtained by rescaling model M1La for $z=2$.
The disk scale-length $r_h$ is only $76$ pc for GR8 ,
while those of the other models are, respectively,  $1.3$ or $2$ kpc (HSBs)
and $1.5$, $3.2$ or $4.8$ kpc (LSBs). The total mass-to-light ratios 
at $3r_h$ (close to the peak of the rotation curves) are, respectively, 
$6$ (HSBs), $12$  (LSBs) and $32$(GR8).

\section{Evolution of dwarf galaxies}

As the Milky Way halo is modeled as an external potential, 
dynamical friction is neglected, which is a good approximation
for satellites $\sim 100$ times less massive than the primary halo
(Colpi et al. 1999). Satellites start at the virial radius of the
primary (their apocenter) as if they were infalling for the first
time, moving on orbits whose ratio between apocenter and pericenter
ranges from 4 to 10, in agreement with simulations of galaxy and cluster
formation (Ghigna 1998). 
Orbital periods are typically of the order of $3-4$ Gyr, but are as
short as $1-2$ Gyr in the run with GR8, because this model was evolved in a
Milky Way potential scaled down in size and mass as expected at $z=2$.  The
inclination and spin of the disk relative to the orbital plane are
randomly selected. In total 40 different runs were performed.

\subsection{Dynamical Evolution}

As the dwarfs approach pericenter (typically of
40-70 kpc) LSBs lose most of
their dark matter halo and stars, due to their low density halos
and large disks, and become weakly bar unstable.  HSBs
suffer modest stripping and their more self-gravitating disks develop
a strong bar. 
The GR8 model is barely stripped owing to its very
dense halo and its small disk radius. Minimal stripping keeps its
disk more self gravitating compared to the structurally similar
LSB and thus a fairly strong bar can develop after the second 
pericenter passage.
Mass stripping for the various models is reflected in the evolution 
of their circular velocity profiles (Figure 1).

After completion of 2-3 orbits in 7 Gyr, our ``tidally stirred'' dwarf
galaxies bear a striking resemblance to the real dSphs
(right panel in Figure 2): direct tidal
heating coupled with the buckling of the bar due to bending
instabilities (Raha et al. 1991) transmute the small disks into
spheroids supported by velocity dispersion instead of rotation (Figure 3).
The GR8-like dwarf, falling into the Milky Way halo around redshift 2,
suffers several (5) strong tidal shocks by the present time and is
thus transformed despite being extremely compact.

\medskip
\epsfxsize=8truecm
\epsfbox{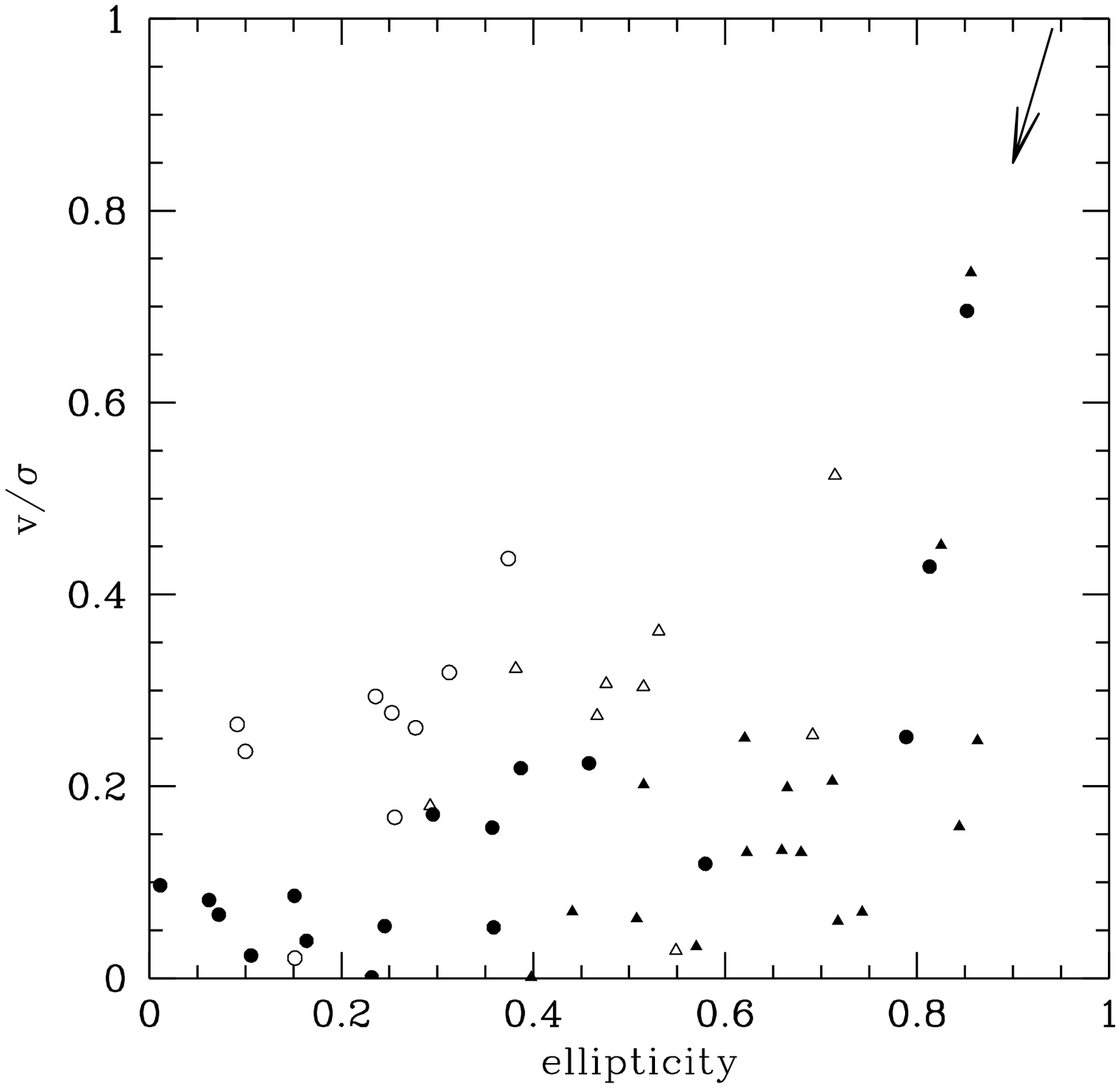}
\figcaption[vsigmaapjl.ps]{\label{fig:asymptotic}    
\small{The final $v/\sigma$ normalized to the initial value is plotted
against the final ellipticity for 26 different runs. For each final
state we show the measurements for a line-of-sight aligned with the major
axis of the remnant (circles) and with the intermediate axis of the
remnant (triangles). Minor axis projections are not shown because rotation
along the other two axes is negligible. Filled symbols are for LSBs, open
symbols are for HSBs. The arrow starts at the coordinates of the initial
states of all the satellites and shows the direction of evolution.
The ellipticity is either is $1-c/b$ or $1-c/a$ (with $a$,$b$ and $c$
being, respectively, the major,intermediate and minor axis of the remnants)
depending on the projection.}}
\medskip

On average the final $v/\sigma$ (i.e. the ratio of the rotational to
random velocity) inside the half mass radius $R_e$ drops  mean
values $\leq 0.5$ and $\sigma$ varies in the
range $7-35$ km s$^{-1}$, as observed for dSphs and dEs in the LG (Ma98).  
The largest of
these values are typical of HSB remnants  and are comparable to the velocity
dispersion
measured for the dEs associated to M31 (Ma98; Kormendy 1987).
Only when the satellite is initially in retrograde
rotation with respect to its orbital motion we measure a final
$v/\sigma$ still close to 1.
The surface brightness profiles remain close to exponential (Figure
1), although with a smaller scale-length 
(typically by a factor of $\sim 2$). The remnants of HSBs 
exhibit a more pronounced steepening 
of the profile inside $R_e$ because stars more efficiently lose
angular momentum due to the strong bar instability. A
From this set of more than 40 runs a clear trend emerges:
{\it LSBs evolve into objects resembling dSphs while HSBs transmute
into dEs}.

\subsection{Gasdynamics and star formation history}

Dwarf irregular galaxies have in general extended gaseous
disks with an average total HI-to-stellar mass ratio larger than 
one (Hoffman et al. 1996): however, within the optical radius 
($\sim$ 3 $r_h$), the neutral hydrogen fraction often drops to 
only $50\%$ of the stellar mass (Jobin \& Carignan 1990; Cote et al. 1990).
In some of our model galaxies we include a
gaseous disk of 20,000 particles which only extends out to the radius
of the stellar disk (material lying outside this radius is 
entirely stripped according to the collisionless runs)
and whose mass is 30\% of the total baryonic mass
(the gas density drops to zero at a radius $R < 0.5r_{h}$
to mimic the ``holes'' found in many dIrrs).  The dynamics
of gas are implemented using an SPH scheme and radiative cooling for a
primordial mixture of hydrogen and helium
(Wadsley et al. 2000).  The initial temperature
of the gas is set at 5000 K.

\medskip
\epsfxsize=8truecm
\epsfbox{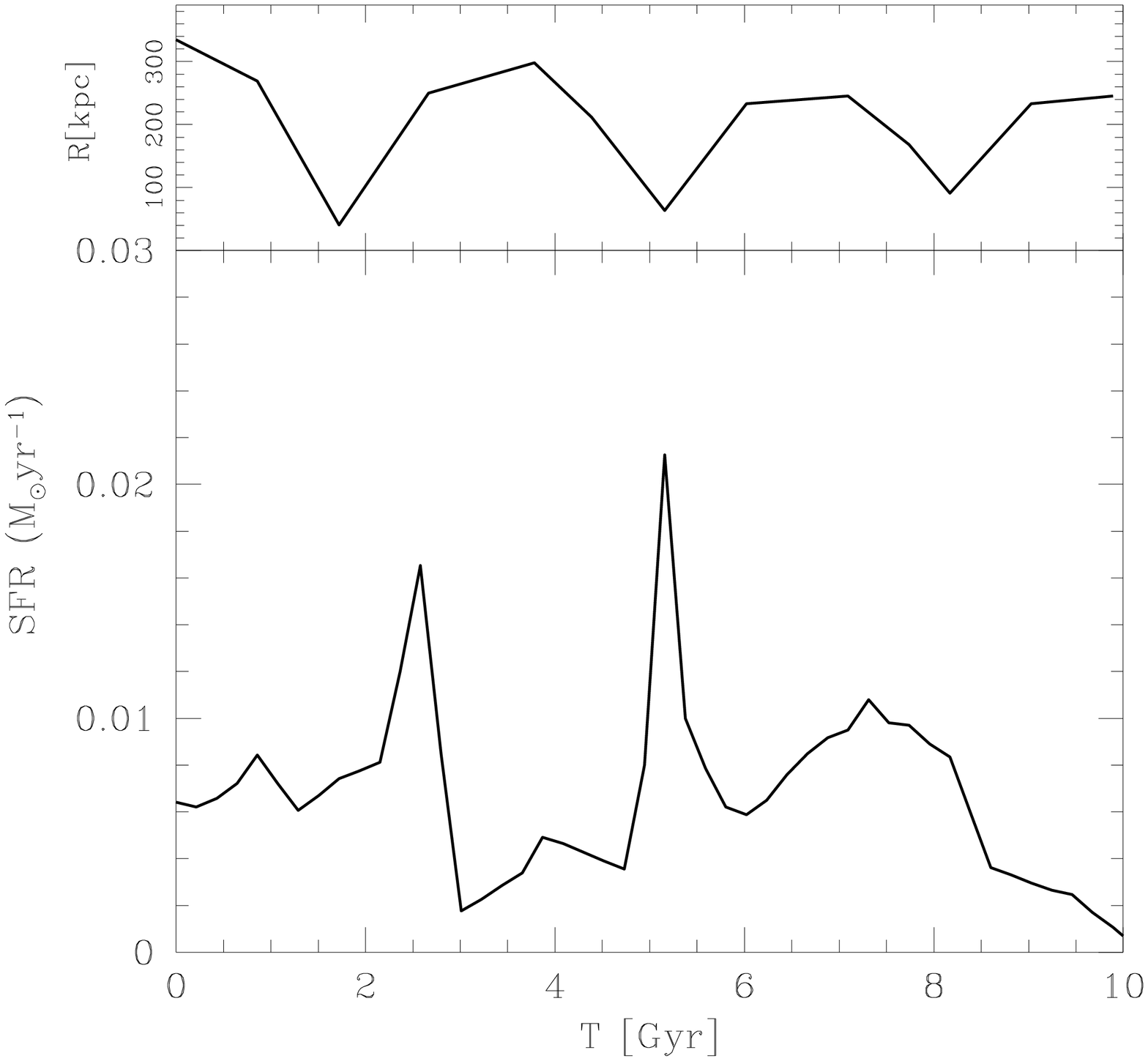}
\figcaption[sfhistnewLSB.ps]{\label{fig:asymptotic}  
\small{Star formation history for an LSB dwarf
We assume that the dwarf enters the Milky Way halo 10 Gyr ago 
(t=0 is the present time). The evolution of the radial orbital
oscillation is also shown in the small panel on top.}}
\medskip

We place an LSB satellite (model M2La) on a 9:1 orbit: 
$\sim 50\%$ of the gas is stripped after two pericenter
passages and is never reaccreted, while the rest is torqued by the
weak bar and gradually flows to the center: the surface density
profile of the gas bound to the system 
steepens remarkably at each pericenter passage 
due to tidal compression and torques.  
We then use the Kennicutt's law (Kennicutt
1998) to determine the star formation rate from the gas surface density.
We also take into account the reduction of the gas mass as it is converted
into stars. The resulting star formation history 
has two main peaks roughly separated by the orbital time
of the dwarf  ($\sim 3.5$ Gyr) (Figure 4), as recently found
for Leo I and Carina (Hernandez et al. 2000).
After 10 Gyr the star formation is suppressed because of gas consumption.
The stronger bar instability in an HSB (model M2H)
placed on the same orbit funnels
more than $80 \%$ of the gas to the center at the first pericentric
passage, giving rise to a starburst ten times stronger than
the bursts in the LSB and using up all the gas in  $\sim 2$ Gyr.
As the strength of the bar instability seems to determine the
type of star formation history, including gas in the GR8 model would
lead to a result qualitatively similar to that of the HSB.
Interestingly, in the LG  both the dEs and 
the extreme dSphs like Draco and Ursa Minor formed the bulk of their 
stellar population during a single early episode (Ma98: Gr99).

Finally we convolve the star formation history with
the passive luminosity evolution of the stellar component
resulting from population synthesis models (Bruzual \& Charlot 1993) 
for low-metallicity systems (1/4 of the solar value).
The multiple bursts are modeled as decreasing exponential
laws with amplitude and time constants constrained by the numerical results.
The resulting total B band luminosities and stellar mass-to-light ratios
of the final remnants are in good agreement with those of observed dSphs. 

\medskip

\section{Discussion}

Figure 5 summarizes the main observable properties of the simulated
satellites projecting them on the Fundamental Plane (FB94).
The remnants of LSB satellites resemble dSphs like
Fornax or Sagittarius ($-14 < M_B < -11$), while HSBs transform in the
bright dEs ($M_{B} > -17$), having a final central surface brightness
higher than that of observed dIrrs with the same luminosities and
therefore matching another observational constraint (FB94;Ma98).
The total (including dark matter) final mass-to-light  ratios
are in the range $6-20$.
Remarkably, our model can reproduce the properties of even the most
extreme dSphs, Draco and Ursa Minor.  In fact,  as the
dark matter halo of GR8 is barely affected by tides
(Figure 1), the remnant ($M_B =
-7.5$) has a final mass-to-light ratio still $\sim 50$ and the central
dark matter density is still around $0.3 M_{\odot}$ pc$^{-3}$,
matching the structural parameters inferred for Draco and Ursa Minor
(Ma98).

\medskip
\epsfxsize=8truecm
\epsfbox{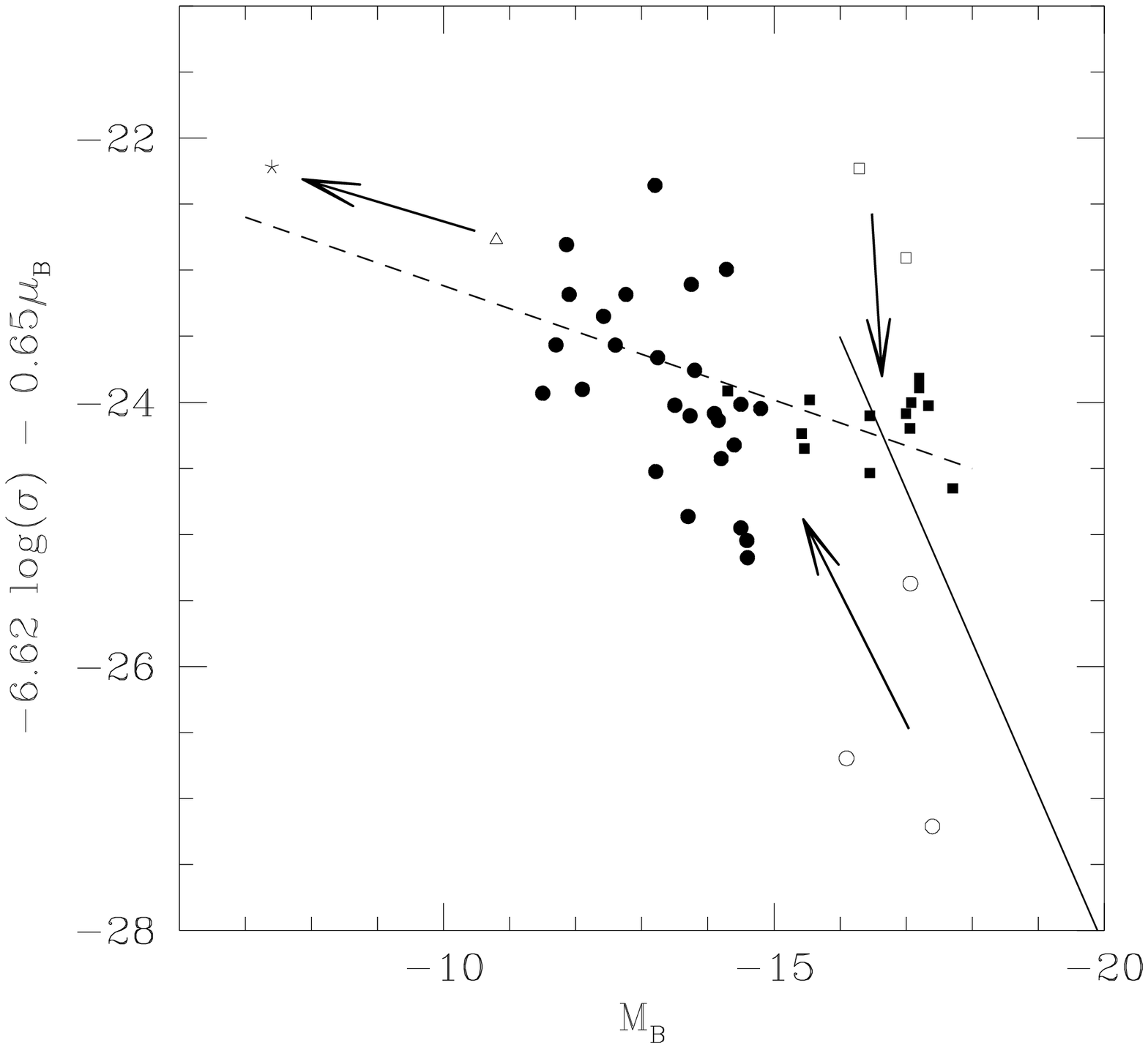}
\figcaption[fundpnew.ps]{\label{fig:asymptotic}  
\small{Fundamental Plane (FP) for all the remnants
as as in Ferguson \& Binggeli (1994) (bottom).
In the FP plot, $\mu$ and  $\sigma$ are, respectively, the
average surface brightness and velocity dispersion 
measured inside $R_e$ and $M_B$ is measured at the Holmberg radius.
The dashed line is a fit to the
distribution of local dSphs, while
the solid line refers 
to elliptical galaxies within the
Virgo cluster (Dressler et al. 1987).
Open and filled symbols represent, respectively, 
the initial and final state of
HSBs (squares) and LSBs (circles).
The initial and final state of the GR8 model are indicated by
the open triangle and ``star''.}}

``Tidal stirring'' naturally leads to the spatial
segregation of dIrrs versus dSph as its effectiveness depends 
strongly on
the distance from the primary.  How important is our
assumption of a massive and extended dark matter halo surrounding the
Milky Way ? When we adopt a ``minimal'' dark halo truncated at 50 kpc
(with mass $5 \times 10^{11} M_{\odot}$; Little \& Tremaine 1987),
tides are too weak and the final remnants are still rotationally
flattened ($v/\sigma > 1$).  Instead, within a halo as massive and
extended as implied by theories of galaxy formation (Peebles 1989) ,
our dIrrs models transform into dSphs even on orbits with apocenters
larger than 200 kpc, explaining the origin of even the farthest dSphs
as Leo I and Leo II.

Though rather speculative at this stage, it is tempting to relate 
HSB satellites observed during the strong bursting phase to
the population of blue compact dwarfs identified by Guzm\'an et al. (1997) at
intermediate redshift. Redshift surveys will establish if bursting dwarfs
have nearby massive companions.

Extended tidal
streams of stars originate from our simulated dwarfs (Fig. 2) 
with a maximum surface brightness of just 30 mag
arcsec$^{-2}$ (B band).  Spectroscopic evidence for stellar streams
from the dSph Carina has been recently claimed (Majewski et al. 2000).
Future astrometric missions, like SIM and GAIA (Gilmore et al. 1998; Helmi et 
should reveal such faint features and will also carry out high-quality
measurements of proper motions for many satellites of the Milky Way,
thus providing a test for the orbital configurations used in this
model.

Our model successfully explains the origin of dSphs 
once all observational constraints are taken into account:
they evolved from dIrrs that entered the halo
of the Milky Way or M31 several Gyr ago moving on plunging
orbits and  suffered stirring by the tidal field of the 
large spirals.

\vskip 16 pt

The authors thank G.Bothun for stimulating discussions.
Simulations were carried out
at the CINECA (Bologna) and ARSC (Fairbanks) supercomputing centers.

\end{document}